\documentclass[twocolumn]{aastex63}

\usepackage[figuresright]{rotating}

\received{***, ****}
\revised{***, ****}
\accepted{***, ****}
\submitjournal{ApJ}

\begin{document}
\title{Composition Comparison between ICMEs from Active Regions and Quiet-Sun Regions}

\correspondingauthor{Hongqiang Song}
\email{hqsong@sdu.edu.cn}

\author{Jinrong Li}
\affiliation{Shandong Provincial Key Laboratory of Optical Astronomy and Solar-Terrestrial Environment, and Institute of Space Sciences, Shandong University, Weihai, Shandong 264209, China}

\author{Hongqiang Song}
\affiliation{Shandong Provincial Key Laboratory of Optical Astronomy and Solar-Terrestrial Environment, and Institute of Space Sciences, Shandong University, Weihai, Shandong 264209, China}

\author{Qi Lv}
\affiliation{Shandong Provincial Key Laboratory of Optical Astronomy and Solar-Terrestrial Environment, and Institute of Space Sciences, Shandong University, Weihai, Shandong 264209, China}

\author{Hui Fu}
\affiliation{Shandong Provincial Key Laboratory of Optical Astronomy and Solar-Terrestrial Environment, and Institute of Space Sciences, Shandong University, Weihai, Shandong 264209, China}

\author{Leping Li}
\affiliation{CAS Key Laboratory of Solar Activity, National Astronomical Observatories, Chinese Academy of Sciences, Beijing, 100101, China}

\author{Ruisheng Zheng}
\affiliation{Shandong Provincial Key Laboratory of Optical Astronomy and Solar-Terrestrial Environment, and Institute of Space Sciences, Shandong University, Weihai, Shandong 264209, China}

\author{Yao Chen}
\affiliation{Shandong Provincial Key Laboratory of Optical Astronomy and Solar-Terrestrial Environment, and Institute of Space Sciences, Shandong University, Weihai, Shandong 264209, China}
\affiliation{Institute of Frontier and Interdisciplinary Science, Shandong University, Qingdao, Shandong 266237, China}

\begin{abstract}
The composition, including the ionic charge states and elemental abundances of heavy elements, within interplanetary coronal mass ejections (ICMEs) has tight correlations with their source regions and eruption processes. This can help analyze the eruption mechanisms and plasma origins of CMEs, and deepen our understanding of energetic solar activities. The active regions and quiet-Sun regions have different physical properties, thus from a statistical point of view, ICMEs originating from the two types of regions should exhibit different compositional characteristics. To demonstrate the differences comprehensively, we conduct survey studies on the ionic charge states of five elements (Mg, Fe, Si, C, and O) and the relative abundances of six elements (Mg/O, Fe/O, Si/O, C/O, Ne/O, and He/O) within ICMEs from 1998 February to 2011 August through the data of advanced composition explorer. The results show that ICMEs from active regions have higher ionic charge states and relative abundances than those from quiet-Sun regions. For the active-region ICMEs, we further analyze the relations between their composition and flare class, and find a positive relationship between them, i.e., the higher classes of the associated flares, the larger means of ionic charge states and relative abundances (except the C/O) within ICMEs. As more (less) fractions of ICMEs originate from active regions around solar maximum (minimum), and active-region ICMEs usually are associated with higher-class flares, our studies might answer why ICME composition measured near 1 au exhibits the solar cycle dependence.
\end{abstract}



\section{Introduction}
Interplanetary coronal mass ejections \citep[ICMEs;][]{manchester17} are counterpart of coronal mass ejections \citep[CMEs;][]{forbes00,forbes06}, which result from eruption of magnetic flux ropes \citep[MFRs;][]{chenpengfei11,webb12} that can form prior to \citep{chengxin11,patsourakos13} and during \citep{song14a,ouyang15,wangwensi17} eruptions. ICMEs can cause extreme space weather effects and affect human high-tech activities \citep{gosling91,zhangjie07,xumengjiao19}. Investigating CMEs and ICMEs will help us predict the space weather, and make timely precautions to reduce losses.

The composition of ICMEs, referring to the ionic charge states and relative abundances of heavy elements, can be obtained through in situ measurements directly. As the composition does not alter during ICME propagation in the interplanetary space \citep[e.g.,][]{owens18}, ICMEs at 1 au and near the Sun have the identical charge states and elemental abundances, which reflect the eruption characteristics and plasma origins of CMEs. Therefore, the in situ composition of ICMEs opens an important avenue for investigating CMEs (see a recent review by \cite{song20b}).

According to the CME models \citep[e.g.,][]{linjun00}, plasma heated by magnetic reconnections in the current sheet is transformed into the outer shell of MFRs alone with the reconnected magnetic field lines \citep{song15a,song16,yejing21}. As CMEs propagate outward, their electron density decreases rapidly. The ionization and recombination processes will be shut down when the density is low enough, making the ionic charge states to freeze-in \citep{owocki83}. Therefore, the ionic charge states near 1 au can be used to infer the information of electron temperature, density, as well as the ICME speed near the Sun \citep{landi12}, see also \cite{shimijie19} for discussions in other contexts. For the metallic elements, such as Mg, Fe and Si, their average values of ionic charge states, i.e., the ${<Q_{Mg}>}$, ${<Q_{Fe}>}$, and ${<Q_{Si}>}$, are sensitive to the electron temperature \citep{lepri01,lepri04,lepri13}. For the non-metallic elements, such as C and O, the charge state ratios, e.g., C$^{6+}$/C$^{5+}$, C$^{6+}$/C$^{4+}$ and O$^{7+}$/O$^{6+}$, are used to indicate the temperature of coronal sources when freezing-in \citep{zhaoliang14,zhaoliang17a}.

The elemental abundances of ICMEs are also employed to diagnose some issues related to CMEs, such as the plasma origin \citep{song17a}, due to the first ionization potential (FIP) effect. The FIP effect is an elemental fractionation that occurs between the solar photosphere and the corona. The low-FIP elements (e.g., Mg, Fe, Si) are enhanced in abundance when they flow into the corona compared to the high-FIP elements (e.g., O). This leads to the relative abundances such as Fe/O higher in the corona \citep{laming04,laming09,laming15}. Therefore, previous studies explore the sources of ICME plasmas through comparing their elemental abundances with photospherical values. For example, \cite{lepri21} reported that the elemental abundances of prominences are similar to the photospheric values, which does not support the prominence material originating from the condensation of coronal plasma.

Many survey studies on the ICME composition have been conducted in recent years. \cite{zurbuchen16} reproted that the ICMEs with elevated Fe charge states have higher FIP fractionation than the other ICMEs. \cite{owens18} and \cite{huangjin20} found that fast ICMEs possess higher charge states and relative abundances compared to slow ones. \cite{guchaoran20} and \cite{song21a} demonstrated that the ICME composition exhibits an obvious solar cycle dependence. Further, \cite{song22a} compared the He abundance between ICMEs and solar wind, and concluded that both ICMEs and slow solar wind possess the solar cycle dependence, while the fast solar wind does not show the dependence.

As both ICMEs and slow wind can originate from active regions and quiet-Sun regions, and more active regions appear in the solar atmosphere at solar maximum, it is natural that more fractions of ICMEs and slow wind detected near 1 au originate from active regions around maximum \citep[e.g.,][]{fuhui17}. If there exist obvious compositional differences between ICMEs from active regions and quiet-Sun regions, the solar cycle dependence of ICME composition can be explained reasonably by means of their source regions. Therefore, a comparative study on the ICME compositions between active regions and quiet-Sun regions is necessary. That is the major motivation for us to conduct this study. This paper is organized as follows. In Section 2, we introduce the instruments used in this study. In Section 3, the observations and results are displayed, which is followed by a summary and discussion in the final section.

\section{Instruments}\label{sec2}
The in-situ compositional data of ICMEs used in this study are provided by the Solar Wind Ion Composition Spectrometer \citep[SWICS;][]{gloeckler98} aboard the Advanced Composition Explorer (ACE) located at the Lagrangian L1 point between the Sun and the Earth. SWICS performs electrostatic selection, time-of-flight analysis, and total energy measurements for each solar wind ion. Through this triple combination measurement, SWICS can determine the mass, charge, and energy of solar wind ions of He, C, N, O, Ne, Mg, Si, S, and Fe, in the energy-per-charge range of 0.49--60 keV e$^{-1}$. The SWICS data used in this paper have been recalibrated with an improved algorithm \citep{shearer14}, which resolves ion species with greater accuracy and includes charge state distributions with uncertainties less than 25\%. The temporal resolutions are 1 and 2 hrs for ionic charge states and elemental abundances, respectively.

The corresponding CMEs and their source regions are examined through the Large Angle and Spectrometric Coronagraph \citep[LASCO;][]{brueckner95} and the Extreme Ultraviolet Imaging Telescope \citep[EIT;][]{delaboudiniere95} on board the Solar and Heliospheric Observatory. The soft X-ray (1--8 \AA) fluxes of associated flares are from the Geostationary Operational Environment Satellite (GOES) that provides the integrated X-ray emission of the whole solar disk.

\section{Observations and Results}\label{sec3}

\subsection{ICMEs and Classifications}
This study uses the online ICME catalog at the ACE science center\footnote{http://www.srl.caltech.edu/ACE/ASC/level2/index.html} \citep{richardson10}, as this catalog is based on ACE measurements. The optimal SWICS data are available from 1998 February to 2011 August, during which 319 ICMEs in total are listed in the catalog. For some ICMEs, the catalog also provides their corresponding CME information near the Sun, i.e., builds connections between ICMEs at L1 point and corresponding CMEs near the Sun. In total 146 ICMEs have their corresponding information of CMEs observed by the LASCO.

Employing the online catalog of LASCO CMEs \citep{gopalswamy09}, we first inspect the animations of LASCO and EIT, and try to find the source regions for each CME manually based on the temporal and spatial correlations. Then the ICMEs are classified into three groups according to their source regions. In the 146 ICMEs with CME information, we find the source regions for 96 cases eventually, including 82 ones originating from active regions (Group I), 11 from quiet-Sun regions (Group II) and 3 from the intermediate regions (Group III). Group III means that the source regions of CMEs cover both the active and quiet-Sun regions. In order to make the statistical results fully represent the differences between ICMEs from the active regions and quiet-Sun regions, the 3 ICMEs in Group III are not included in this survey study. In the meantime, we also examine the GOES soft X-ray fluxes and find the associated flares for 72 active-region CMEs based on the GOES flare list$\footnote{https://www.ngdc.noaa.gov/stp/space-weather/solar-data/solar-features/solarflares/x-rays/goes/xrs/}$.

The information for the 96 ICMEs with identified source regions is listed in Table 1. The first column gives the event number. Columns 2--4 show the disturbance time (typically related to the arrival of the ICME shock at Earth), start and end times of the ICME based on plasma and magnetic field observations. Column 5 is the time of the associated CME first appearance in the field of view of LASCO. The next two columns display the CME group and the NOAA active-region numbers, and the last column shows the flare class associated with the CME. The active-region numbers are acquired from the website of Solar Monitor$\footnote{www.solarmonitor.org}$, and flare information, from the GOES flare list.

\startlongtable
\centerwidetable
\begin{deluxetable*}{cccccccc}
	\tabletypesize{\small}
	\tablewidth{0pt}
	\tablecaption{The information of 96 ICMEs (CMEs) with identified source regions. Note that the Universal Time is used. \label{Table 1}}
	\tablehead{
		\colhead{No.} & \colhead{Disturbance Start} & \colhead{ICME Start} & \colhead{ICME End}& \colhead{CME in LASCO} & \colhead{Group} &\colhead{AR No.}& \colhead{Flare Class}
	}\startdata
	1 & 1998 May 01 21:56 & May 02 05:00 & May 04 02:00 & 1998 Apr 29 16:58 & I &  08210 & M 6.8  \\
	2 & 1998 Nov 07 08:15 & Nov 07 22:00 & Nov 09 01:00 & 1998 Nov 04 07:54 & I &  08375 & C 1.6  \\
	3 & 1998 Nov 08 04:51 & Nov 09 01:00 & Nov 11 01:00 & 1998 Nov 05 20:44 & I &  08375 & M 8.4  \\
	4 & 1998 Nov 13 01:43 & Nov 13 02:00 & Nov 14 12:00 & 1998 Nov 09 18:18 & I &  08378 & ...   \\
	5 & 1999 Apr 16 11:25 & Apr 16 18:00 & Apr 17 19:00 & 1999 Apr 13 03:30 & II & ... & ...   \\
	6 & 1999 Jun 26 20:16 & Jun 27 22:00 & Jun 29 04:00 & 1999 Jun 24 13:31 & I &  08595 & C 4.1  \\
	7 & 1999 Jul 06 15:09 & Jul 06 21:00 & Jul 07 02:00 & 1999 Jul 03 19:54 & I &  08610 & ...  \\
	8 & 1999 Jul 26 23:33 & Jul 27 17:00 & Jul 29 12:00 & 1999 Jul 23 21:30 & I &  08636 & ...  \\
	9 & 1999 Jul 31 18:37 & Jul 31 19:00 & Aug 02 06:00 & 1999 Jul 28 09:06 & I &  08649 & M 2.3  \\
	10 & 1999 Aug 11 23:00 & Aug 12 03:00 & Aug 14 00:00 & 1999 Aug 09 03:26 & I &  08657 & C 4.3  \\
	11 & 1999 Aug 20 23:00 & Aug 20 23:00 & Aug 23 11:00 & 1999 Aug 17 13:31 & III & ... & ...  \\
	12 & 1999 Sep 22 12:22 & Sep 22 19:00 & Sep 24 03:00 & 1999 Sep 20 06:06 & II & ... & ...  \\
	13 & 1999 Oct 21 02:25 & Oct 21 08:00 & Oct 22 07:00 & 1999 Oct 18 00:06 & II & ... & ...  \\
	14 & 2000 Jun 22 00:23 & Jun 22 17:00 & Jun 23 02:00 & 2000 Jun 18 17:54 & I &  08831 & M 3.9  \\
	15 & 2000 Feb 11 02:58 & Feb 11 16:00 & Feb 11 20:00 & 2000 Feb 08 09:30 & I &  08858 & M 1.3  \\
	16 & 2000 Feb 11 23:52 & Feb 12 12:00 & Feb 13 00:00 & 2000 Feb 10 02:30 & I &  08858 & C 7.3  \\
	17 & 2000 Feb 14 07:31 & Feb 14 12:00 & Feb 16 08:00 & 2000 Feb 12 04:31 & I & 08858 & M 1.7  \\
	18 & 2000 Feb 20 21:39 & Feb 21 05:00 & Feb 22 12:00 & 2000 Feb 17 20:06 & I &  08872 & M 2.5  \\
	19 & 2000 Apr 06 16:39 & Apr 07 06:00 & Apr 08 06:00 & 2000 Apr 04 16:32 & I &  08933 & C 9.7  \\
	20 & 2000 Jun 04 15:02 & Jun 04 22:00 & Jun 06 22:00 & 2000 May 31 08:06 & II & ... & ...  \\
	21 & 2000 Jun 08 09:10 & Jun 08 12:00 & Jun 10 17:00 & 2000 Jun 06 15:54 & I &  09026 & X 2.3  \\
	22 & 2000 Jul 15 14:37 & Jul 15 19:00 & Jul 17 08:00 & 2000 Jul 14 10:54 & I &  09077 & X 5.7  \\
	23 & 2000 Jul 26 18:57 & Jul 27 02:00 & Jul 28 02:00 & 2000 Jul 23 05:30 & III & ... & ...  \\
	24 & 2000 Jul 28 06:34 & Jul 28 12:00 & Jul 30 13:00 & 2000 Jul 25 03:30 & I &  09097 & M 8.0  \\
	25 & 2000 Aug 11 18:45 & Aug 12 05:00 & Aug 13 22:00 & 2000 Aug 09 16:30 & I &  09114 & C 2.3  \\
	26 & 2000 Sep 08 12:00 & Sep 08 12:00 & Sep 10 10:00 & 2000 Sep 05 05:54 & I &  09152 & ...  \\
	27 & 2000 Oct 05 03:26 & Oct 05 13:00 & Oct 07 11:00 & 2000 Oct 02 20:26 & I &  09176 & ...  \\
	28 & 2000 Oct 28 09:54 & Oct 28 21:00 & Oct 29 22:00 & 2000 Oct 25 08:26 & I &  09199 & ...  \\
	29 & 2000 Dec 22 19:25 & Dec 23 00:00 & Dec 23 12:00 & 2000 Dec 18 11:50 & I &  09269 & C 7.0  \\
	30 & 2001 Mar 03 11:21 & Mar 04 04:00 & Mar 05 02:00 & 2001 Feb 28 14:50 & I &  09364 & ...  \\
	31 & 2001 Mar 27 17:47 & Mar 28 17:00 & Mar 30 18:00 & 2001 Mar 25 17:06 & I &  09402 & C 9.0  \\
	32 & 2001 Mar 31 00:52 & Mar 31 05:00 & Mar 31 22:00 & 2001 Mar 28 12:50 & I &  09397 & ...  \\
	33 & 2001 Mar 31 22:00 & Apr 01 04:00 & Apr 03 15:00 & 2001 Mar 29 10:26 & I &  09393 & X 1.7  \\
	34 & 2001 Apr 04 14:55 & Apr 04 18:00 & Apr 05 12:00 & 2001 Apr 02 22:06 & I &  09393 & X 20.0  \\
	35 & 2001 Apr 08 11:01 & Apr 08 14:00 & Apr 09 04:00 & 2001 Apr 06 19:30 & I &  09415 & X 5.6  \\
	36 & 2001 Apr 11 13:43 & Apr 11 22:00 & Apr 13 07:00 & 2001 Apr 10 05:30 & I &  09415 & X 2.3  \\
	37 & 2001 Apr 13 07:34 & Apr 13 09:00 & Apr 14 12:00 & 2001 Apr 11 13:31 & I &  09415 & M 2.3  \\
	38 & 2001 Apr 28 05:01 & Apr 28 14:00 & May 01 02:00 & 2001 Apr 26 12:30 & I &  09433 & M 7.8  \\
	39 & 2001 Sep 30 19:24 & Oct 01 08:00 & Oct 02 00:00 & 2001 Sep 28 08:54 & I &  09636 & M 3.3  \\
	40 & 2001 Oct 01 21:15 & Oct 02 04:00 & Oct 02 12:00 & 2001 Sep 29 11:54 & I &  09636 & M 1.8  \\
	41 & 2001 Oct 11 17:01 & Oct 12 04:00 & Oct 12 09:00 & 2001 Oct 09 11:30 & I &  09653 & M 1.4  \\
	42 & 2001 Oct 21 16:48 & Oct 21 20:00 & Oct 25 10:00 & 2001 Oct 19 16:50 & I &  09661 & X 1.6  \\
	43 & 2001 Oct 26 22:00 & Oct 27 03:00 & Oct 28 12:00 & 2001 Oct 22 18:26 & I &  09672 & X 1.2  \\
	44 & 2001 Oct 28 03:19 & Oct 29 22:00 & Oct 31 13:00 & 2001 Oct 25 15:26 & I &  09672 & X 1.3  \\
	45 & 2001 Nov 19 18:15 & Nov 19 22:00 & Nov 21 13:00 & 2001 Nov 17 05:30 & III & ... & ...  \\
	46 & 2001 Nov 24 06:56 & Nov 24 14:00 & Nov 25 20:00 & 2001 Nov 22 23:30 & I &  09704 & M 9.9  \\
	47 & 2002 Mar 18 13:22 & Mar 19 05:00 & Mar 20 16:00 & 2002 Mar 15 23:06 & I &  09866 & M 2.2  \\
	48 & 2002 Apr 17 11:07 & Apr 17 16:00 & Apr 19 15:00 & 2002 Apr 15 03:50 & I &  09906 & M 1.2  \\
	49 & 2002 Apr 19 08:35 & Apr 20 00:00 & Apr 21 18:00 & 2002 Apr 17 08:26 & I &  09906 & M 2.6  \\
	50 & 2002 May 11 10:14 & May 11 15:00 & May 12 00:00 & 2002 May 08 13:50 & I &  09934 & C 4.2  \\
	51 & 2002 May 20 03:40 & May 20 10:00 & May 21 22:00 & 2002 May 16 00:50 & I &  09948 & C 4.5  \\
	52 & 2002 May 23 10:50 & May 23 20:00 & May 25 18:00 & 2002 May 22 03:26 & II & ... & ...  \\
	53 & 2002 Jul 17 16:03 & Jul 18 12:00 & Jul 19 09:00 & 2002 Jul 15 20:30 & I &  10030 & X 3.0  \\
	54 & 2002 Aug 18 18:46 & Aug 19 12:00 & Aug 21 14:00 & 2002 Aug 16 12:30 & I &  10069 & M 5.2  \\
	55 & 2002 Sep 07 16:36 & Sep 08 04:00 & Sep 08 20:00 & 2002 Sep 05 16:54 & I &  10102 & C 5.2  \\
	56 & 2002 Sep 19 06:00 & Sep 19 20:00 & Sep 20 21:00 & 2002 Sep 17 08:06 & I &  10114 & C 8.6  \\
	57 & 2003 Feb 01 13:05 & Feb 01 19:00 & Feb 03 07:00 & 2003 Jun 30 10:06 & II & ... & ...  \\
	58 & 2003 Jun 16 18:00 & Jun 17 10:00 & Jun 18 08:00 & 2003 Jun 14 01:54 & II & ... & ...  \\
	59 & 2003 Oct 24 15:24 & Oct 24 21:00 & Oct 25 12:00 & 2003 Oct 22 08:30 & I &  10484 & ...  \\
	60 & 2003 Oct 28 02:06 & Oct 28 02:30 & Oct 28 09:00 & 2003 Oct 26 17:54 & I &  10484 & X 1.2  \\
	61 & 2003 Oct 29 06:11 & Oct 29 11:00 & Oct 30 03:00 & 2003 Oct 28 11:30 & I &  10486 & X 17.2  \\
	62 & 2003 Oct 30 16:19 & Oct 31 02:00 & Nov 02 00:00 & 2003 Oct 29 20:54 & I &  10486 & X 10.0  \\
	63 & 2003 Nov 20 08:03 & Nov 20 10:00 & Nov 21 08:00 & 2003 Nov 18 08:50 & I &  10501 & M 3.9  \\
	64 & 2004 Jun 22 01:37 & Jun 22 08:00 & Jun 23 17:00 & 2004 Jun 20 00:06 & I &  10540 & ...  \\
	65 & 2004 Jun 23 14:25 & Jun 23 23:00 & Jun 25 04:00 & 2004 Jun 21 04:54 & II & ... & ...  \\
	66 & 2004 Jul 22 10:36 & Jul 22 18:00 & Jul 24 08:00 & 2004 Jul 20 13:31 & I &  10652 & M 8.6  \\
	67 & 2004 Jul 24 06:13 & Jul 24 14:00 & Jul 25 15:00 & 2004 Jul 22 07:31 & I &  10652 & C 2.0  \\
	68 & 2004 Jul 25 15:00 & Jul 25 20:00 & Jul 26 22:00 & 2004 Jul 23 16:06 & I &  10652 & C 1.0  \\
	69 & 2004 Jul 26 22:49 & Jul 27 02:00 & Jul 27 22:00 & 2004 Jul 25 14:54 & I &  10652 & M 2.2  \\
	70 & 2004 Nov 07 18:27 & Nov 07 22:00 & Nov 09 10:00 & 2004 Nov 04 23:30 & I &  10696 & M 5.4  \\
	71 & 2004 Nov 09 18:25 & Nov 09 20:00 & Nov 11 23:00 & 2004 Nov 07 16:54 & I &  10696 & X 2.0  \\
	72 & 2004 Nov 11 17:10 & Nov 12 08:00 & Nov 13 23:00 & 2004 Nov 10 02:26 & I &  10696 & X 2.5  \\
	73 & 2004 Dec 11 13:40 & Dec 12 22:00 & Dec 13 19:00 & 2004 Dec 08 20:26 & I &  10709 & C 2.5  \\
	74 & 2005 Jun 08 17:00 & Jun 08 21:00 & Jun 09 18:00 & 2005 Jun 05 15:30 & II & ... & ...  \\
	75 & 2005 Jun 16 11:00 & Jun 16 14:00 & Jun 17 07:00 & 2005 Jun 13 17:54 & I &  10718 & C 4.2  \\
	76 & 2005 Jun 18 21:00 & Jun 18 23:00 & Jun 20 03:00 & 2005 Jun 17 09:30 & I &  10720 & X 3.8  \\
	77 & 2005 Jun 21 17:11 & Jun 21 19:00 & Jun 22 17:00 & 2005 Jun 20 06:54 & I &  10720 & X 7.1  \\
	78 & 2005 Feb 20 12:00 & Feb 20 12:00 & Feb 22 07:00 & 2005 Feb 17 00:06 & I &  10734 & C 4.9  \\
	79 & 2005 May 15 02:38 & May 15 06:00 & May 19 00:00 & 2005 May 13 17:12 & I &  10759 & M 8.0 \\
	80 & 2005 May 20 03:00 & May 20 03:00 & May 22 02:00 & 2005 May 16 13:50 & I &  10759 & C 1.2  \\
	81 & 2005 Jul 10 03:37 & Jul 10 10:00 & Jul 12 04:00 & 2005 Jul 07 17:06 & I &  10786 & M 4.9  \\
	82 & 2005 Aug 09 00:00 & Aug 09 00:00 & Aug 09 19:00 & 2005 Aug 05 08:54 & I &  10792 & C 2.6  \\
	83 & 2005 Aug 24 06:13 & Aug 24 14:00 & Aug 24 23:00 & 2005 Aug 22 01:31 & I &  10798 & M 2.6  \\
	84 & 2005 Sep 02 14:19 & Sep 02 18:00 & Sep 03 04:00 & 2005 Aug 31 11:30 & I &  10803 & C 2.0  \\
	85 & 2006 Jul 09 21:36 & Jul 10 21:00 & Jul 11 19:00 & 2006 Jul 06 08:54 & I &  10898 & M 2.5  \\
	86 & 2006 Aug 19 11:31 & Aug 20 13:00 & Aug 21 16:00 & 2006 Aug 16 16:30 & I &  10904 & C 3.6  \\
	87 & 2006 Dec 14 14:14 & Dec 14 22:00 & Dec 15 13:00 & 2006 Dec 13 02:54 & I &  10930 & X 3.4  \\
	88 & 2006 Dec 16 17:55 & Dec 17 00:00 & Dec 17 17:00 & 2006 Dec 14 22:30 & I &  10930 & X 1.5  \\
	89 & 2008 Dec 16 11:59 & Dec 17 03:00 & Dec 17 14:00 & 2008 Dec 12 08:54 & II & ... & ...  \\
	90 & 2010 Feb 11 00:00 & Feb 11 08:00 & Feb 12 03:00 & 2010 Feb 07 03:54 & I &  11045 & M 6.4  \\
	91 & 2010 Apr 05 08:26 & Apr 05 12:00 & Apr 06 14:00 & 2010 Apr 03 10:33 & I &  11059 & B 7.4  \\
	92 & 2010 Apr 11 13:04 & Apr 12 01:00 & Apr 12 15:00 & 2010 Apr 08 04:54 & I &  11060 & B 1.5  \\
	93 & 2010 May 28 02:58 & May 28 19:00 & May 29 17:00 & 2010 May 24 14:06 & II & ... & ...  \\
	94 & 2011 Feb 18 01:30 & Feb 18 19:00 & Feb 20 08:00 & 2011 Feb 15 02:36 & I &  11158 & X 2.2  \\
	95 & 2011 Aug 04 21:53 & Aug 05 05:00 & Aug 05 14:00 & 2011 Aug 02 06:36 & I &  11261 & M 1.4  \\
	96 & 2011 Aug 05 17:51 & Aug 06 22:00 & Aug 07 22:00 & 2011 Aug 04 04:12 & I &  11261 & M 9.3  \\
	\enddata
		
\end{deluxetable*}
\vspace{-0.5cm}

\subsection{Comparison between ICMEs from different regions}
Figure 1 shows the histograms of ionic charge states for ICMEs from active regions (red) and quiet-Sun regions (blue). From Panel (a) to (f), the ${<Q_{Mg}>}$, ${<Q_{Fe}>}$, ${<Q_{Si}>}$, C$^{6+}$/C$^{4+}$, C$^{6+}$/C$^{5+}$, and O$^{7+}$/O$^{6+}$ are displayed sequentially. Based on all of the corresponding data points we calculate the average values for each parameter, which are 9.424 (9.277), 12.510 (11.531), 10.279 (10.006), 9.630 (5.401), 2.055 (1.302), and 0.956 (0.606) correspondingly for ICMEs from active regions (quiet-Sun regions). These values are also marked with arrows in each panel. Overall, all of the ionic charge states are higher in ICMEs from active regions compared to those from quiet-Sun regions. This agrees with the expectation as the temperatures of reconnection regions are higher for active-region CMEs.

The charge state histograms of metallic elements (${<Q_{Mg}>}$, ${<Q_{Fe}>}$ and ${<Q_{Si}>}$) are obviously different for the two groups of ICMEs. The active-region ICMEs contain higher charge states. For example, the active-region ICMEs include a fraction of Fe ions with charge states higher than 16+, while the quiet-region ICMEs does not at all. For the non-metallic elements, the average values of C$^{6+}$/C$^{4+}$, C$^{6+}$/C$^{5+}$, and O$^{7+}$/O$^{6+}$ within ICMEs from active regions are obviously higher than those from quiet-Sun regions by 78.3$\%$, 57.8$\%$, and 57.8$\%$, respectively.

Figure 2 shows the histograms of relative elemental abundances for ICMEs from active regions (red) and quiet-Sun regions (blue). The six panels display the Mg/O, Fe/O, Si/O, C/O, Ne/O, and He/O sequentially, also with the corresponding average values being displayed in each panel. It is obvious that the elemental abundances of active-region ICMEs contain substantial fraction of higher abundances, and the average values of each parameter for active-region ICMEs are higher than the quiet-region ICMEs by 59.8$\%$ (from 0.184 to 0.294), 61.1$\%$ (from 0.185 to 0.298), 44.1$\%$ (from 0.179 to 0.258), 10.5$\%$ (from 0.486 to 0.537), 56.9$\%$ (from 0.167 to 0.262) and 71.7$\%$ (from 78.418 to 134.677), respectively. Note that large overlaps exist between the composition histograms of active-region and quiet-region ICMEs, although there are significant differences.

\subsection{Comparison between ICMEs with different flare classes}
The comparison between ICMEs from different regions clearly demonstrate that the active-region ICMEs have elevated ionic charge states and enriched elemental abundances, which should correlate with the high temperatures of reconnection regions and magnetic structures of active regions. To further confirm this speculation, we examine the correlations between ICME composition and flare classes. As mentioned, we find the associated flares for 72 active-region ICMEs, including 21 (29, 20) events associated with C-class (M-class, X-class) flares, except 2 ICMEs associated with B-class flares, which are not considered here due to limited number.

Figure 3 presents the histograms of ${<Q_{Mg}>}$, ${<Q_{Fe}>}$, ${<Q_{Si}>}$, C$^{6+}$/C$^{4+}$, C$^{6+}$/C$^{5+}$ and O$^{7+}$/O$^{6+}$ for ICMEs associated with C-class (blue), M-class (orange) and X-class (red) flares. The arrows and numbers in each panel represent the corresponding means. To demonstrate the variations between ICMEs with weak and strong flares, we focus on the values of C-class and X-class flares. The average values for C-class flares are 9.323, 11.951, 10.062, 8.217, 1.832, and 0.679, respectively, while for X-class flares they are 9.570, 13.296, 10.557, 11.988, 2.322, and 1.319 correspondingly. It is clear that the ionic charge states of ICMEs associated with X-class flares are higher than those associated with C-class flares, agreeing with our expectation.

Figure 4 displays the histograms of Mg/O, Fe/O, Si/O, C/O, Ne/O and He/O for ICMEs associated with C-class (blue), M-class (orange) and X-class (red) flares. Also the arrows and numbers represent the corresponding means in each panel. The average values of each parameter associated with C-class (X-class) flares are 0.255 (0.336), 0.245 (0.320), 0.236 (0.276), 0.534 (0.536), 0.199 (0.388) and 122.649 (147.362), respectively. Again it is clear that the ICMEs associated with X-class flares possess higher elemental abundances compared to those associated with C-class flares. The statistical results in Figures 3 and 4 support that ICMEs associated with stronger flares possess more elevated ionic charge states and enriched elemental abundances. As both the M-class and X-class flares belong to strong events, their histograms of composition are relatively similar.

In addition, we perform a t-test using the Python routine $stats.ttest\_ind$ to examine whether the average values of two groups are significantly different. The function runs the independent sample t-test and outputs a p-value and the test-statistic. The p-value ranges from 0.0 to 1.0, and a small value (typically less than 0.05 or 0.01) indicates that the averages of the two groups are truly different. The t-test is conducted on all compositional parameters in Figures 1--4, i.e., between ICMEs from active regions and quiet-Sun regions, as well as ICMEs associated with C-class and X-class flares. The results show that almost all of the p-values are less than 0.01 except that of the C/O of ICMEs associated with C-class and X-class flares, whose p-value is 0.928. This illustrates that most averages of the compositional parameters are significantly different at the 99\% confidence level.

\section{Summary and Discussion}\label{sec4}
Based on the ICME catalog available at the ACE science center \citep{richardson10} and the CME catalog at the CDAW \citep{gopalswamy09}, we manually found the source regions for 96 ICMEs from 1998 to 2011, including 82 ones from active regions, 11 from quiet-Sun regions, and 3 from the intermediate regions. The composition comparison between active-region and quiet-region ICMEs showed that the events from active regions possess elevated ionic charge states and enriched elemental abundances compared to those from quiet-Sun regions. For the 82 active-region ICMEs, we found the corresponding flares for 72 of them, including 21 (29, 20) events associated with C-class (M-class, X-class) flares, except 2 events associated with B-class flares. The comparison between ICMEs with different flare classes demonstrated that ICMEs associated with X-class flares exhibited higher charge states and elemental abundances (except the C/O) compared to C-class flares.

Our survey studies support that the elevated charge states within ICMEs mainly result from the higher temperatures of reconnection regions during solar eruptions \citep{lepri04,song16}, and agree with the expectations of popular CME models \citep{mikic94,antiochos99,linjun00,moore01,torok04,kliem06,chenpengfei08}. Our studies also support that the elemental abundances within ICMEs depend on both the magnetic structures of source regions and the X-ray features of associated flares, which are mainly correlated with the FIP effect \citep{laming04,laming15} and photoionization \citep{shemi91,schmelz93}, respectively. More discussions can be found in \cite{song21a} and \cite{song22a}.

As mentioned previous studies showed that the composition of ICMEs \citep{guchaoran20,song21a,song22a} and slow solar wind \citep{aellig01,kasper07,kasper12,McIntosh11,lepri13,alterman19,alterman21} have obvious solar cycle dependence. The solar cycle is well represented by the sunspot number, and the number of active regions increases from solar minimum to maximum. Therefore, it is natural that more CMEs occur in active regions around solar maximum, which leads to more fractions of ICMEs detected near 1 au originate from active regions and are associated with stronger flares. This might answer why ICME composition measured near 1 au exhibits the solar cycle dependence \citep{song21a,song22a}.

The situation is similar for slow solar wind, which can emanate from both active regions and quiet-Sun regions \citep[e.g.,][]{zhaoliang17a}. More fractions of slow wind emanate from active (queit-Sun) regions around solar maximum (minimum) \citep[e.g.,][]{fuhui17}, thus slow wind composition measured near 1 au also exhibits the solar cycle dependence \citep{lepri13,song22a}. On the contrary, the fast solar wind always originates from coronal holes, which leads to their composition, such as the He abundance, detected near 1 au does not exhibit variation as obvious as the slow wind \citep{alterman19,alterman21,song22a} and ICMEs \citep{song22a} during the solar cycle.

The FIP effect is a main elemental abundance feature of the slow wind and ICMEs, while the gravitational settling \citep{raymond97,raymond98,lenz98,raymond99} can modify signatures of the FIP effect in both the slow wind \citep{weberg12} and ICMEs \citep{rivera22}. \cite{rivera22} reported that the ICMEs between 1998 and 2011 exhibit some gravitational settling effects in $\sim$33\% of the ICME periods. They also found that the effect is most prominent during solar minimum. The gravitational settling induces mass-proportional depletion of elements in the corona, and leads to the decrease of relative abundances such as Fe/O. This could also be a factor responsible for the solar cycle dependence of the relative abundances of partial heavy elements \citep{song21a}.

The histograms of Mg/O, Fe/O, and Si/O in Figure 2 support the idea of strong FIP enhancement in active regions, while the Ne/O and He/O distributions do not as the FIPs of Ne (21.56 eV) and He (24.58 eV) are higher than that of O (13.61 eV). \cite{song21a} suggested that the enhanced Ne/O in ICMEs from active regions could result from the photoionization effect of flares \citep{shemi91,schmelz93}. The pre-flare soft X-ray can create a slab-like region of non-thermal ionization ratios at the chromosphere base \citep{shemi91}. As the photoionization cross section ratio of Ne and O is 9:4 \citep{yeh85} and the photoionized O can recombine through the charge transfer reaction with the neutral H, the Ne ions are mixed with the neutral O in the slab region and selected with the thermally ionized low-FIP elements for preferential transfer to the solar corona and ICMEs \citep{song21a}.

Besides, there exist bare ion anomalies in both slow wind \citep{zhaoliang17b,raymond22} and ICMEs \citep{kocher17}, in which the C$^{6+}$ and other fully stripped ions are unusually low and they are called outlier wind or ICMEs. \cite{rivera21} reported that the Ne/O is enhanced in the outlier ICMEs, which can also contribute to the high Ne/O of ICMEs from active regions. The number of outlier events is positively correlated with the solar cycle, so it could be another factor responsible for the solar cycle dependence of the relative abundances of some elements \citep{song21a}.

However, both the photoionization effect and outlier can not explain the high He/O in active-region ICMEs, because O has larger photoionization cross section than He \citep{yeh85} and the unusually low He$^{2+}$ density of outlier ICMEs reduces the He/O (not shown). Since the mass of O is heavier than that of He, the gravitational settling might be adopted to explain the high He/O in active regions. More studies are necessary to clarify this speculation.

\acknowledgments We thank the referee (Dr. John Raymond) for his helpful comments and suggestions. Hongqiang Song thanks Drs. Xin Cheng and Bo Li for their discussions. We acknowledge the use of the ICME catalog at the ACE science center \citep{richardson10} and the CME catalog at the CDAW \citep{gopalswamy09}. All the SWICS data are downloaded from the ACE science center. This work is supported by the NSFC grants U2031109, 11973031, 12073042, and the National Key R\&D Program of China 2022YFF0503003 (2022YFF0503000). Hongqiang Song is also supported by the CAS grants XDA-17040507.

\begin{figure*}[htb!]
\epsscale{0.9} \plotone{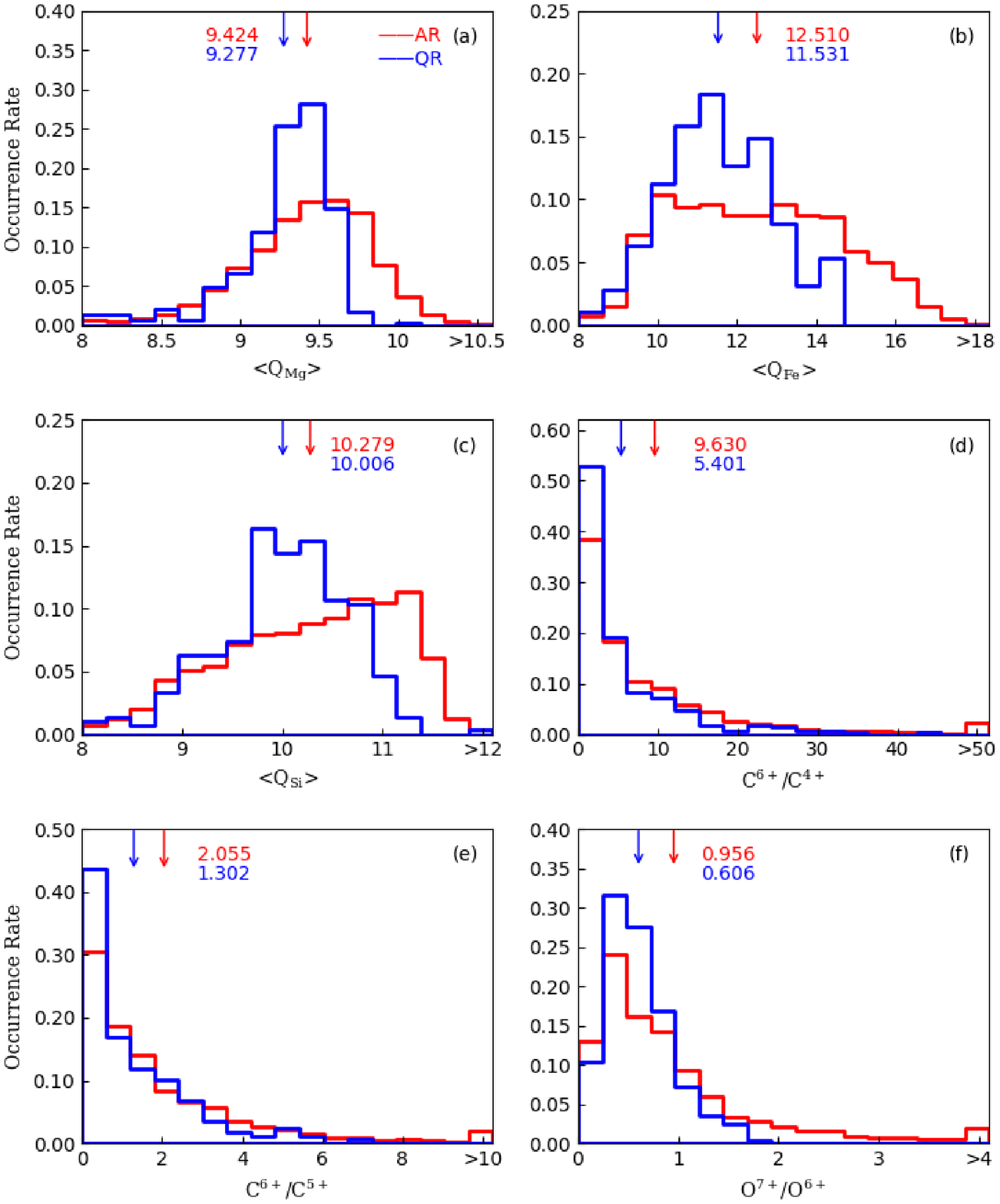} \caption{Histograms of the ${<Q_{Mg}>}$ (a), ${<Q_{Fe}>}$ (b), ${<Q_{Si}>}$ (c), C$^{6+}$/C$^{4+}$ (d), C$^{6+}$/C$^{5+}$ (e) and O$^{7+}$/O$^{6+}$ (f) in the ICMEs originating from active regions (red) and quiet-Sun regions (blue). The arrows and numbers in each panel indicate the corresponding average values. \label{Figure 1}}
\end{figure*}

\begin{figure*}[htb!]
\epsscale{0.9} \plotone{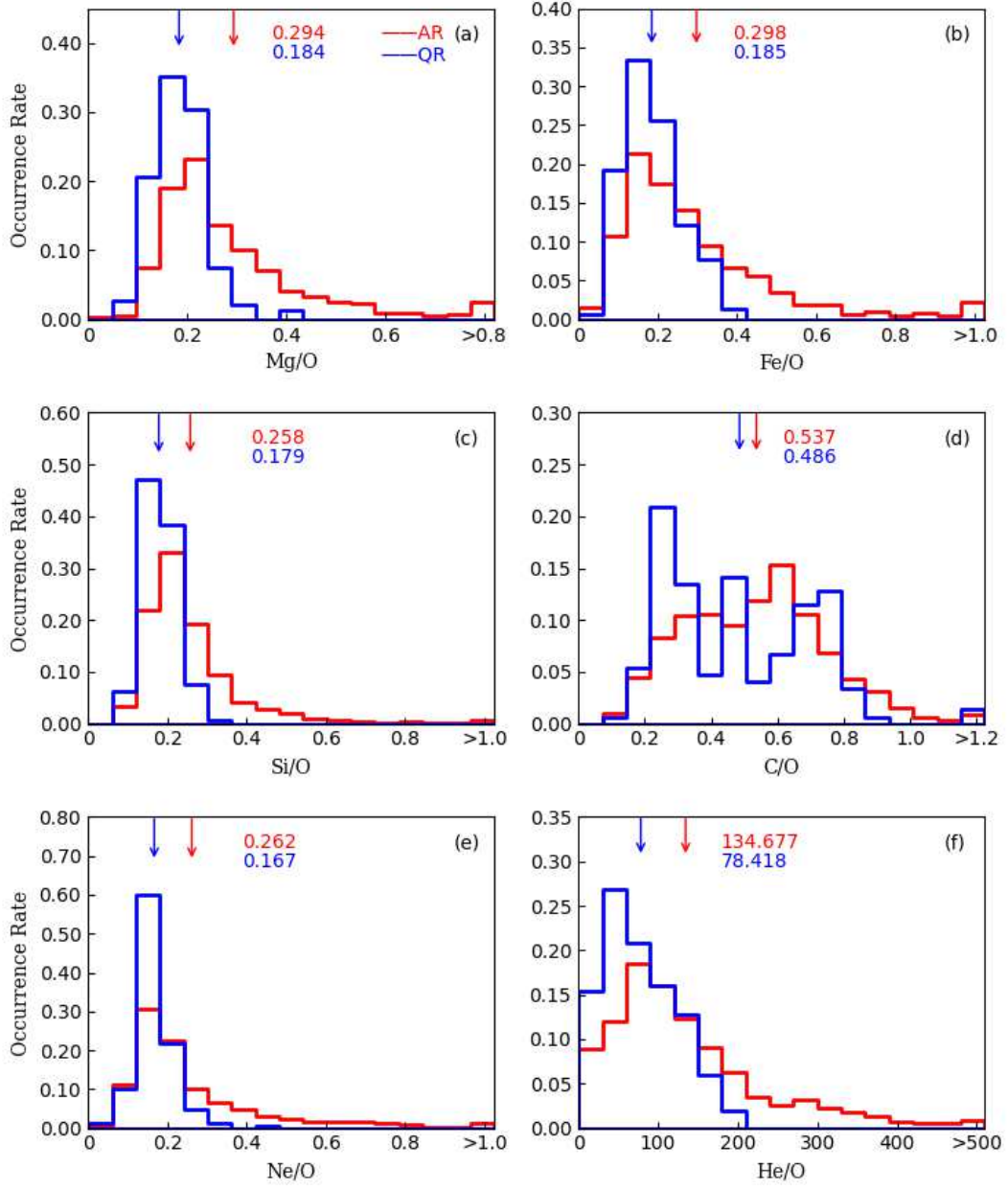} \caption{Histograms of the Mg/O (a), Fe/O (b), Si/O (c), C/O (d), Ne/O (e) and He/O (f) in the ICMEs originating from active regions (red) and quiet-Sun regions (blue). The arrows and numbers in each panel indicate the corresponding average values. \label{Figure 2}}
\end{figure*}

\begin{figure*}[htb!]
\epsscale{0.9} \plotone{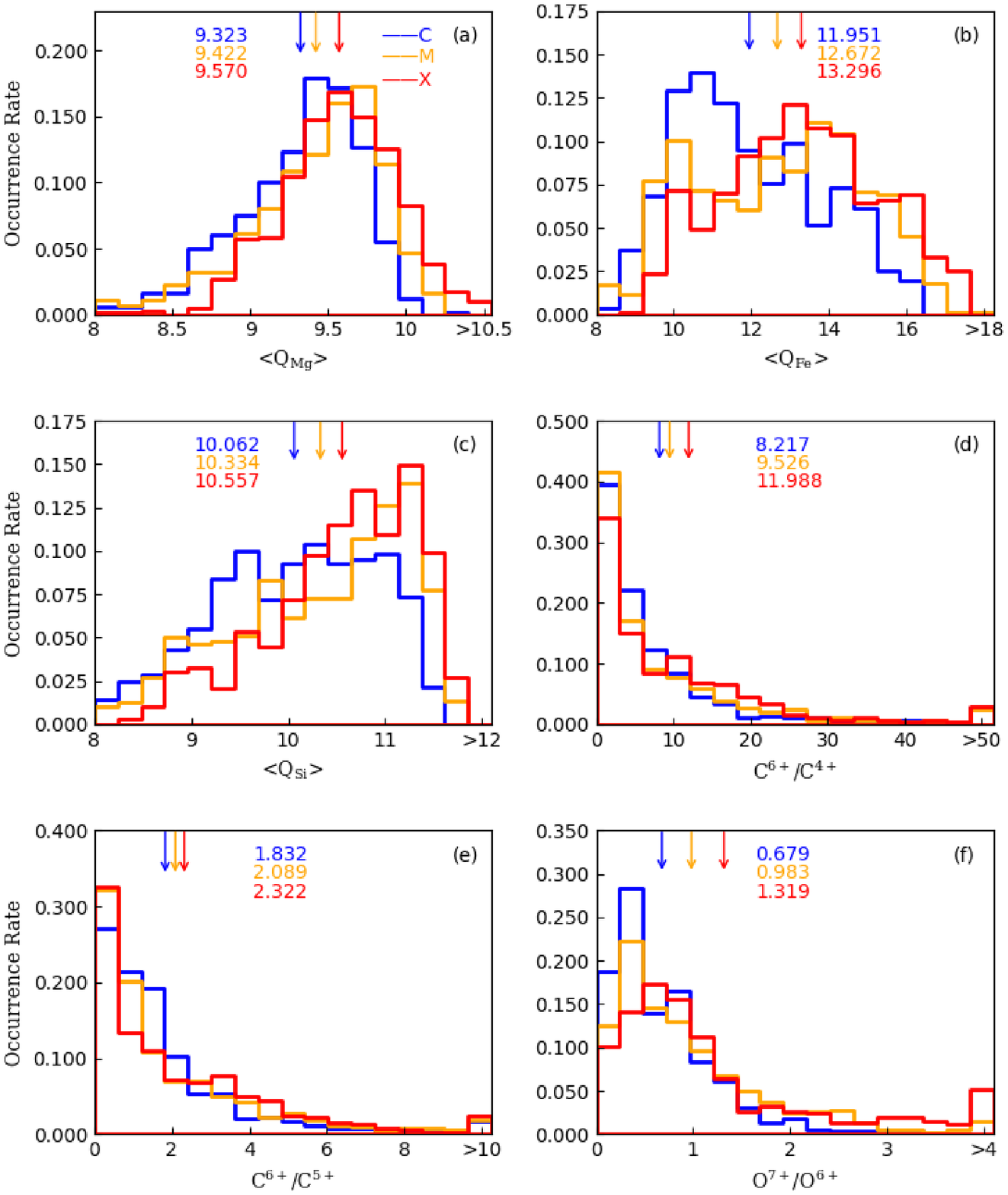} \caption{Histograms of the ${<Q_{Mg}>}$ (a), ${<Q_{Fe}>}$ (b), ${<Q_{Si}>}$ (c), C$^{6+}$/C$^{4+}$ (d), C$^{6+}$/C$^{5+}$ (e) and O$^{7+}$/O$^{6+}$ (f) in the ICMEs associated with C-class (blue), M-class (orange), and X-class (red) flares. The arrows and numbers in each panel indicate the corresponding average values. \label{Figure 3}}
\end{figure*}

\begin{figure*}[htb!]
\epsscale{0.9} \plotone{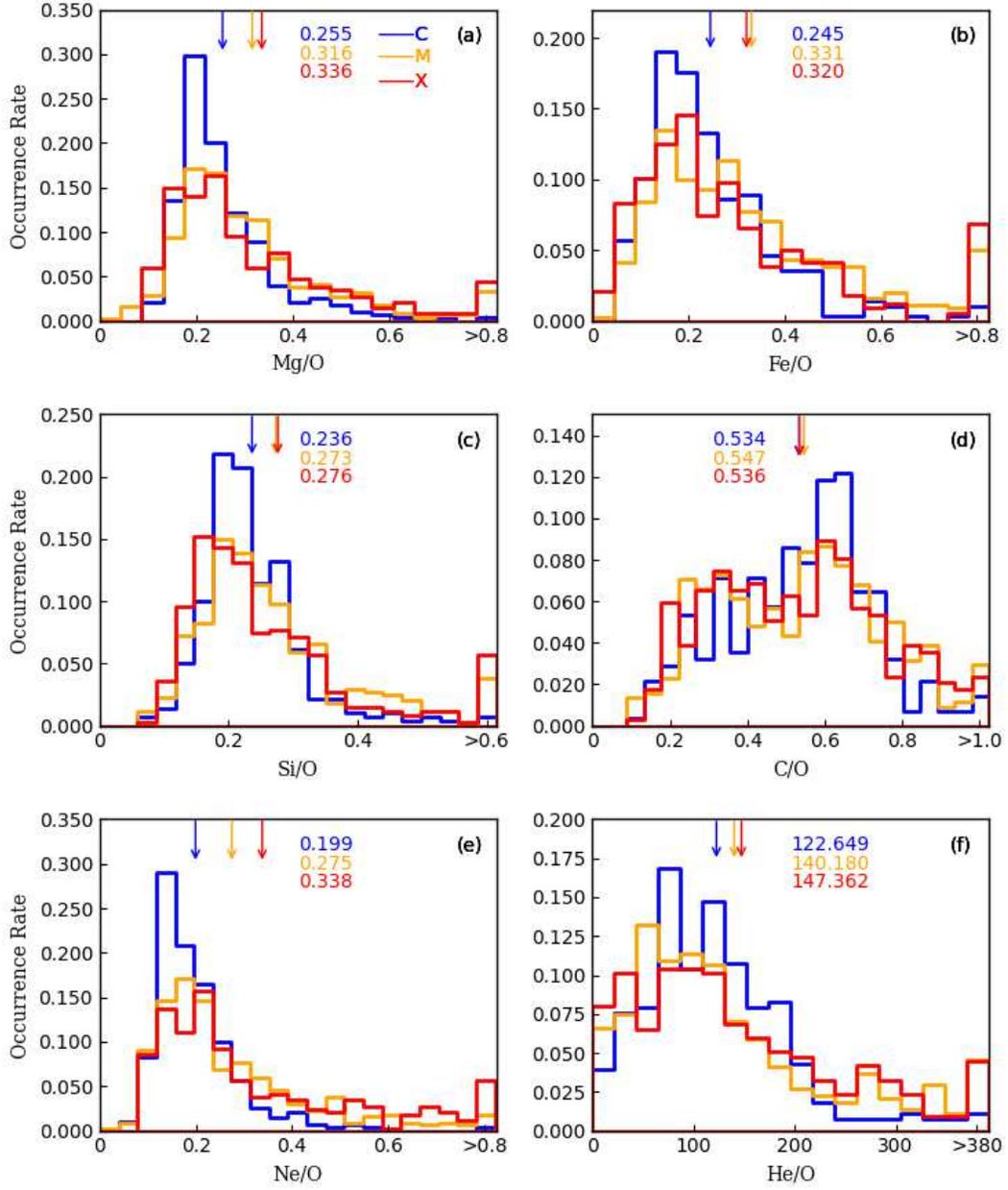} \caption{Histograms of the Mg/O (a), Fe/O (b), Si/O (c), C/O (d), Ne/O (e) and He/O (f) in the ICMEs associated with C-class (blue), M-class (orange), and X-class (red) flares. The arrows and numbers in each panel indicate the corresponding average values. \label{Figure 4}}
\end{figure*}


\begin{thebibliography}{}
\expandafter\ifx\csname natexlab\endcsname\relax\def\natexlab#1{#1}\fi
\providecommand{\url}[1]{\href{#1}{#1}}
\providecommand{\dodoi}[1]{doi:~\href{http://doi.org/#1}{\nolinkurl{#1}}}
\providecommand{\doeprint}[1]{\href{http://ascl.net/#1}{\nolinkurl{http://ascl.net/#1}}}
\providecommand{\doarXiv}[1]{\href{https://arxiv.org/abs/#1}{\nolinkurl{https://arxiv.org/abs/#1}}}

\bibitem[{{Aellig} {et~al.}(2001){Aellig}, {Lazarus}, \&
  {Steinberg}}]{aellig01}
{Aellig}, M.~R., {Lazarus}, A.~J., \& {Steinberg}, J.~T. 2001, \grl, 28, 2767,
  \dodoi{10.1029/2000GL012771}

\bibitem[{{Alterman} \& {Kasper}(2019)}]{alterman19}
{Alterman}, B.~L., \& {Kasper}, J.~C. 2019, \apjl, 879, L6,
  \dodoi{10.3847/2041-8213/ab2391}

\bibitem[{{Alterman} {et~al.}(2021){Alterman}, {Kasper}, {Leamon}, \&
  {McIntosh}}]{alterman21}
{Alterman}, B.~L., {Kasper}, J.~C., {Leamon}, R.~J., \& {McIntosh}, S.~W. 2021,
  \solphys, 296, 67, \dodoi{10.1007/s11207-021-01801-9}

\bibitem[{{Antiochos} {et~al.}(1999){Antiochos}, {DeVore}, \&
  {Klimchuk}}]{antiochos99}
{Antiochos}, S.~K., {DeVore}, C.~R., \& {Klimchuk}, J.~A. 1999, \apj, 510, 485,
  \dodoi{10.1086/306563}

\bibitem[{{Brueckner} {et~al.}(1995){Brueckner}, {Howard}, {Koomen},
  {Korendyke}, {Michels}, {Moses}, {Socker}, {Dere}, {Lamy}, {Llebaria},
  {Bout}, {Schwenn}, {Simnett}, {Bedford}, \& {Eyles}}]{brueckner95}
{Brueckner}, G.~E., {Howard}, R.~A., {Koomen}, M.~J., {et~al.} 1995, \solphys,
  162, 357, \dodoi{10.1007/BF00733434}

\bibitem[{{Chen}(2008)}]{chenpengfei08}
{Chen}, P.~F. 2008, Journal of Astrophysics and Astronomy, 29, 179,
  \dodoi{10.1007/s12036-008-0023-0}

\bibitem[{{Chen}(2011)}]{chenpengfei11}
---. 2011, Living Reviews in Solar Physics, 8, 1, \dodoi{10.12942/lrsp-2011-1}

\bibitem[{{Cheng} {et~al.}(2011){Cheng}, {Zhang}, {Liu}, \&
  {Ding}}]{chengxin11}
{Cheng}, X., {Zhang}, J., {Liu}, Y., \& {Ding}, M.~D. 2011, \apjl, 732, L25,
  \dodoi{10.1088/2041-8205/732/2/L25}

\bibitem[{{Delaboudini{\`e}re} {et~al.}(1995){Delaboudini{\`e}re}, {Artzner},
  {Brunaud}, {Gabriel}, {Hochedez}, {Millier}, {Song}, {Au}, {Dere}, {Howard},
  {Kreplin}, {Michels}, {Moses}, {Defise}, {Jamar}, {Rochus}, {Chauvineau},
  {Marioge}, {Catura}, {Lemen}, {Shing}, {Stern}, {Gurman}, {Neupert},
  {Maucherat}, {Clette}, {Cugnon}, \& {Van Dessel}}]{delaboudiniere95}
{Delaboudini{\`e}re}, J.~P., {Artzner}, G.~E., {Brunaud}, J., {et~al.} 1995,
  \solphys, 162, 291, \dodoi{10.1007/BF00733432}

\bibitem[{{Forbes}(2000)}]{forbes00}
{Forbes}, T.~G. 2000, \jgr, 105, 23153, \dodoi{10.1029/2000JA000005}

\bibitem[{{Forbes} {et~al.}(2006){Forbes}, {Linker}, {Chen}, {Cid}, {K{\'o}ta},
  {Lee}, {Mann}, {Miki{\'c}}, {Potgieter}, {Schmidt}, {Siscoe}, {Vainio},
  {Antiochos}, \& {Riley}}]{forbes06}
{Forbes}, T.~G., {Linker}, J.~A., {Chen}, J., {et~al.} 2006, \ssr, 123, 251,
  \dodoi{10.1007/s11214-006-9019-8}

\bibitem[{{Fu} {et~al.}(2017){Fu}, {Madjarska}, {Xia}, {Li}, {Huang}, \&
  {Wangguan}}]{fuhui17}
{Fu}, H., {Madjarska}, M.~S., {Xia}, L., {et~al.} 2017, \apj, 836, 169,
  \dodoi{10.3847/1538-4357/aa5cba}

\bibitem[{{Gloeckler} {et~al.}(1998){Gloeckler}, {Cain}, {Ipavich}, {Tums},
  {Bedini}, {Fisk}, {Zurbuchen}, {Bochsler}, {Fischer}, {Wimmer-Schweingruber},
  {Geiss}, \& {Kallenbach}}]{gloeckler98}
{Gloeckler}, G., {Cain}, J., {Ipavich}, F.~M., {et~al.} 1998, \ssr, 86, 497,
  \dodoi{10.1023/A:1005036131689}

\bibitem[{{Gopalswamy} {et~al.}(2009){Gopalswamy}, {Yashiro}, {Michalek},
  {Stenborg}, {Vourlidas}, {Freeland}, \& {Howard}}]{gopalswamy09}
{Gopalswamy}, N., {Yashiro}, S., {Michalek}, G., {et~al.} 2009, Earth Moon and
  Planets, 104, 295, \dodoi{10.1007/s11038-008-9282-7}

\bibitem[{{Gosling} {et~al.}(1991){Gosling}, {McComas}, {Phillips}, \&
  {Bame}}]{gosling91}
{Gosling}, J.~T., {McComas}, D.~J., {Phillips}, J.~L., \& {Bame}, S.~J. 1991,
  \jgr, 96, 7831, \dodoi{10.1029/91JA00316}

\bibitem[{{Gu} {et~al.}(2020){Gu}, {Yao}, \& {Dai}}]{guchaoran20}
{Gu}, C., {Yao}, S., \& {Dai}, L. 2020, \apj, 900, 123,
  \dodoi{10.3847/1538-4357/aba7b8}

\bibitem[{{Huang} {et~al.}(2020){Huang}, {Liu}, {Feng}, {Zhao}, {Abidin},
  {Shen}, \& {Jacob}}]{huangjin20}
{Huang}, J., {Liu}, Y., {Feng}, H., {et~al.} 2020, \apj, 893, 136,
  \dodoi{10.3847/1538-4357/ab7a28}

\bibitem[{{Kasper} {et~al.}(2012){Kasper}, {Stevens}, {Korreck}, {Maruca},
  {Kiefer}, {Schwadron}, \& {Lepri}}]{kasper12}
{Kasper}, J.~C., {Stevens}, M.~L., {Korreck}, K.~E., {et~al.} 2012, \apj, 745,
  162, \dodoi{10.1088/0004-637X/745/2/162}

\bibitem[{{Kasper} {et~al.}(2007){Kasper}, {Stevens}, {Lazarus}, {Steinberg},
  \& {Ogilvie}}]{kasper07}
{Kasper}, J.~C., {Stevens}, M.~L., {Lazarus}, A.~J., {Steinberg}, J.~T., \&
  {Ogilvie}, K.~W. 2007, \apj, 660, 901, \dodoi{10.1086/510842}

\bibitem[{{Kliem} \& {T{\"o}r{\"o}k}(2006)}]{kliem06}
{Kliem}, B., \& {T{\"o}r{\"o}k}, T. 2006, \prl, 96, 255002,
  \dodoi{10.1103/PhysRevLett.96.255002}

\bibitem[{{Kocher} {et~al.}(2017){Kocher}, {Lepri}, {Landi}, {Zhao}, \&
  {Manchester}}]{kocher17}
{Kocher}, M., {Lepri}, S.~T., {Landi}, E., {Zhao}, L., \& {Manchester}, W.~B.,
  I. 2017, \apj, 834, 147, \dodoi{10.3847/1538-4357/834/2/147}

\bibitem[{{Laming}(2004)}]{laming04}
{Laming}, J.~M. 2004, \apj, 614, 1063, \dodoi{10.1086/423780}

\bibitem[{{Laming}(2009)}]{laming09}
---. 2009, \apj, 695, 954, \dodoi{10.1088/0004-637X/695/2/954}

\bibitem[{{Laming}(2015)}]{laming15}
---. 2015, Living Reviews in Solar Physics, 12, 2, \dodoi{10.1007/lrsp-2015-2}

\bibitem[{{Landi} {et~al.}(2012){Landi}, {Gruesbeck}, {Lepri}, {Zurbuchen}, \&
  {Fisk}}]{landi12}
{Landi}, E., {Gruesbeck}, J.~R., {Lepri}, S.~T., {Zurbuchen}, T.~H., \& {Fisk},
  L.~A. 2012, \apj, 761, 48, \dodoi{10.1088/0004-637X/761/1/48}

\bibitem[{{Lenz} {et~al.}(1998){Lenz}, {Lou}, \& {Rosner}}]{lenz98}
{Lenz}, D.~D., {Lou}, Y.-Q., \& {Rosner}, R. 1998, \apj, 504, 1020,
  \dodoi{10.1086/306111}

\bibitem[{{Lepri} {et~al.}(2013){Lepri}, {Landi}, \& {Zurbuchen}}]{lepri13}
{Lepri}, S.~T., {Landi}, E., \& {Zurbuchen}, T.~H. 2013, \apj, 768, 94,
  \dodoi{10.1088/0004-637X/768/1/94}

\bibitem[{{Lepri} \& {Rivera}(2021)}]{lepri21}
{Lepri}, S.~T., \& {Rivera}, Y.~J. 2021, \apj, 912, 51,
  \dodoi{10.3847/1538-4357/abea9f}

\bibitem[{{Lepri} \& {Zurbuchen}(2004)}]{lepri04}
{Lepri}, S.~T., \& {Zurbuchen}, T.~H. 2004, Journal of Geophysical Research
  (Space Physics), 109, A01112, \dodoi{10.1029/2003JA009954}

\bibitem[{{Lepri} {et~al.}(2001){Lepri}, {Zurbuchen}, {Fisk}, {Richardson},
  {Cane}, \& {Gloeckler}}]{lepri01}
{Lepri}, S.~T., {Zurbuchen}, T.~H., {Fisk}, L.~A., {et~al.} 2001, \jgr, 106,
  29231, \dodoi{10.1029/2001JA000014}

\bibitem[{{Lin} \& {Forbes}(2000)}]{linjun00}
{Lin}, J., \& {Forbes}, T.~G. 2000, \jgr, 105, 2375,
  \dodoi{10.1029/1999JA900477}

\bibitem[{{Manchester} {et~al.}(2017){Manchester}, {Kilpua}, {Liu}, {Lugaz},
  {Riley}, {T{\"o}r{\"o}k}, \& {Vr{\v{s}}nak}}]{manchester17}
{Manchester}, W., {Kilpua}, E. K.~J., {Liu}, Y.~D., {et~al.} 2017, \ssr, 212,
  1159, \dodoi{10.1007/s11214-017-0394-0}

\bibitem[{{McIntosh} {et~al.}(2011){McIntosh}, {Kiefer}, {Leamon}, {Kasper}, \&
  {Stevens}}]{McIntosh11}
{McIntosh}, S.~W., {Kiefer}, K.~K., {Leamon}, R.~J., {Kasper}, J.~C., \&
  {Stevens}, M.~L. 2011, \apjl, 740, L23, \dodoi{10.1088/2041-8205/740/1/L23}

\bibitem[{{Mikic} \& {Linker}(1994)}]{mikic94}
{Mikic}, Z., \& {Linker}, J.~A. 1994, \apj, 430, 898, \dodoi{10.1086/174460}

\bibitem[{{Moore} {et~al.}(2001){Moore}, {Sterling}, {Hudson}, \&
  {Lemen}}]{moore01}
{Moore}, R.~L., {Sterling}, A.~C., {Hudson}, H.~S., \& {Lemen}, J.~R. 2001,
  \apj, 552, 833, \dodoi{10.1086/320559}

\bibitem[{{Ouyang} {et~al.}(2015){Ouyang}, {Yang}, \& {Chen}}]{ouyang15}
{Ouyang}, Y., {Yang}, K., \& {Chen}, P.~F. 2015, \apj, 815, 72,
  \dodoi{10.1088/0004-637X/815/1/72}

\bibitem[{{Owens}(2018)}]{owens18}
{Owens}, M.~J. 2018, \solphys, 293, 122, \dodoi{10.1007/s11207-018-1343-0}

\bibitem[{{Owocki} {et~al.}(1983){Owocki}, {Holzer}, \&
  {Hundhausen}}]{owocki83}
{Owocki}, S.~P., {Holzer}, T.~E., \& {Hundhausen}, A.~J. 1983, \apj, 275, 354,
  \dodoi{10.1086/161538}

\bibitem[{{Patsourakos} {et~al.}(2013){Patsourakos}, {Vourlidas}, \&
  {Stenborg}}]{patsourakos13}
{Patsourakos}, S., {Vourlidas}, A., \& {Stenborg}, G. 2013, \apj, 764, 125,
  \dodoi{10.1088/0004-637X/764/2/125}

\bibitem[{{Raymond}(1999)}]{raymond99}
{Raymond}, J.~C. 1999, \ssr, 87, 55, \dodoi{10.1023/A:1005157914229}

\bibitem[{{Raymond} {et~al.}(2022){Raymond}, {Asgari-Targhi}, {Wilson},
  {Rivera}, {Lepri}, \& {Shen}}]{raymond22}
{Raymond}, J.~C., {Asgari-Targhi}, M., {Wilson}, M.~L., {et~al.} 2022, \apj,
  936, 175, \dodoi{10.3847/1538-4357/ac8976}

\bibitem[{{Raymond} {et~al.}(1997){Raymond}, {Kohl}, {Noci}, {Antonucci},
  {Tondello}, {Huber}, {Gardner}, {Nicolosi}, {Fineschi}, {Romoli}, {Spadaro},
  {Siegmund}, {Benna}, {Ciaravella}, {Cranmer}, {Giordano}, {Karovska},
  {Martin}, {Michels}, {Modigliani}, {Naletto}, {Panasyuk}, {Pernechele},
  {Poletto}, {Smith}, {Suleiman}, \& {Strachan}}]{raymond97}
{Raymond}, J.~C., {Kohl}, J.~L., {Noci}, G., {et~al.} 1997, \solphys, 175, 645,
  \dodoi{10.1023/A:1004948423169}

\bibitem[{{Raymond} {et~al.}(1998){Raymond}, {Fineschi}, {Smith}, {Gardner},
  {O'Neal}, {Ciaravella}, {Kohl}, {Marsden}, {Williams}, {Benna}, {Giordano},
  {Noci}, \& {Jewitt}}]{raymond98}
{Raymond}, J.~C., {Fineschi}, S., {Smith}, P.~L., {et~al.} 1998, \apj, 508,
  410, \dodoi{10.1086/306391}

\bibitem[{{Richardson} \& {Cane}(2010)}]{richardson10}
{Richardson}, I.~G., \& {Cane}, H.~V. 2010, \solphys, 264, 189,
  \dodoi{10.1007/s11207-010-9568-6}

\bibitem[{{Rivera} {et~al.}(2021){Rivera}, {Lepri}, {Raymond}, {Reeves},
  {Stevens}, \& {Zhao}}]{rivera21}
{Rivera}, Y.~J., {Lepri}, S.~T., {Raymond}, J.~C., {et~al.} 2021, \apj, 921,
  93, \dodoi{10.3847/1538-4357/ac1676}

\bibitem[{{Rivera} {et~al.}(2022){Rivera}, {Raymond}, {Landi}, {Lepri},
  {Reeves}, {Stevens}, \& {Alterman}}]{rivera22}
{Rivera}, Y.~J., {Raymond}, J.~C., {Landi}, E., {et~al.} 2022, \apj, 936, 83,
  \dodoi{10.3847/1538-4357/ac8873}

\bibitem[{{Schmelz}(1993)}]{schmelz93}
{Schmelz}, J.~T. 1993, \apj, 408, 373, \dodoi{10.1086/172594}

\bibitem[{{Shearer} {et~al.}(2014){Shearer}, {von Steiger}, {Raines}, {Lepri},
  {Thomas}, {Gilbert}, {Landi}, \& {Zurbuchen}}]{shearer14}
{Shearer}, P., {von Steiger}, R., {Raines}, J.~M., {et~al.} 2014, \apj, 789,
  60, \dodoi{10.1088/0004-637X/789/1/60}

\bibitem[{{Shemi}(1991)}]{shemi91}
{Shemi}, A. 1991, \mnras, 251, 221, \dodoi{10.1093/mnras/251.2.221}

\bibitem[{{Shi} {et~al.}(2019){Shi}, {Li}, {Van Doorsselaere}, {Chen}, \&
  {Huang}}]{shimijie19}
{Shi}, M., {Li}, B., {Van Doorsselaere}, T., {Chen}, S.-X., \& {Huang}, Z.
  2019, \apj, 870, 99, \dodoi{10.3847/1538-4357/aaf393}

\bibitem[{{Song} {et~al.}(2022){Song}, {Cheng}, {Li}, {Zhang}, \&
  {Chen}}]{song22a}
{Song}, H., {Cheng}, X., {Li}, L., {Zhang}, J., \& {Chen}, Y. 2022, \apj, 925,
  137, \dodoi{10.3847/1538-4357/ac3bbf}

\bibitem[{{Song} {et~al.}(2021){Song}, {Li}, {Sun}, {Lv}, {Zheng}, \&
  {Chen}}]{song21a}
{Song}, H., {Li}, L., {Sun}, Y., {et~al.} 2021, \solphys, 296, 111,
  \dodoi{10.1007/s11207-021-01852-y}

\bibitem[{{Song} \& {Yao}(2020)}]{song20b}
{Song}, H., \& {Yao}, S. 2020, Sci China Tech Sci, 63, 2171,
  \dodoi{10.1007/s11431-020-1680-y}

\bibitem[{{Song} {et~al.}(2015){Song}, {Chen}, {Zhang}, {Cheng}, {Fu}, \&
  {LI}}]{song15a}
{Song}, H.~Q., {Chen}, Y., {Zhang}, J., {et~al.} 2015, \apjl, 804, L38,
  \dodoi{10.1088/2041-8205/804/2/L38}

\bibitem[{{Song} {et~al.}(2014){Song}, {Zhang}, {Chen}, \& {Cheng}}]{song14a}
{Song}, H.~Q., {Zhang}, J., {Chen}, Y., \& {Cheng}, X. 2014, \apjl, 792, L40,
  \dodoi{10.1088/2041-8205/792/2/L40}

\bibitem[{{Song} {et~al.}(2016){Song}, {Zhong}, {Chen}, {Zhang}, {Cheng},
  {Zhao}, {Hu}, \& {Li}}]{song16}
{Song}, H.~Q., {Zhong}, Z., {Chen}, Y., {et~al.} 2016, \apjs, 224, 27,
  \dodoi{10.3847/0067-0049/224/2/27}

\bibitem[{{Song} {et~al.}(2017){Song}, {Chen}, {Li}, {Li}, {Zhao}, {He},
  {Duan}, {Cheng}, \& {Zhang}}]{song17a}
{Song}, H.~Q., {Chen}, Y., {Li}, B., {et~al.} 2017, \apjl, 836, L11,
  \dodoi{10.3847/2041-8213/aa5d54}

\bibitem[{{T{\"o}r{\"o}k} {et~al.}(2004){T{\"o}r{\"o}k}, {Kliem}, \&
  {Titov}}]{torok04}
{T{\"o}r{\"o}k}, T., {Kliem}, B., \& {Titov}, V.~S. 2004, \aap, 413, L27,
  \dodoi{10.1051/0004-6361:20031691}

\bibitem[{{Wang} {et~al.}(2017){Wang}, {Liu}, {Wang}, {Hu}, {Shen}, {Jiang}, \&
  {Zhu}}]{wangwensi17}
{Wang}, W., {Liu}, R., {Wang}, Y., {et~al.} 2017, Nature Communications, 8,
  1330, \dodoi{10.1038/s41467-017-01207-x}

\bibitem[{{Webb} \& {Howard}(2012)}]{webb12}
{Webb}, D.~F., \& {Howard}, T.~A. 2012, Living Reviews in Solar Physics, 9, 3,
  \dodoi{10.12942/lrsp-2012-3}

\bibitem[{{Weberg} {et~al.}(2012){Weberg}, {Zurbuchen}, \& {Lepri}}]{weberg12}
{Weberg}, M.~J., {Zurbuchen}, T.~H., \& {Lepri}, S.~T. 2012, \apj, 760, 30,
  \dodoi{10.1088/0004-637X/760/1/30}

\bibitem[{{Xu} {et~al.}(2019){Xu}, {Shen}, {Wang}, {Luo}, \&
  {Chi}}]{xumengjiao19}
{Xu}, M., {Shen}, C., {Wang}, Y., {Luo}, B., \& {Chi}, Y. 2019, \apjl, 884,
  L30, \dodoi{10.3847/2041-8213/ab4717}

\bibitem[{{Ye} {et~al.}(2021){Ye}, {Cai}, {Shen}, {Raymond}, {Mei}, {Li}, \&
  {Lin}}]{yejing21}
{Ye}, J., {Cai}, Q., {Shen}, C., {et~al.} 2021, \apj, 909, 45,
  \dodoi{10.3847/1538-4357/abdeb5}

\bibitem[{{Yeh} \& {Lindau}(1985)}]{yeh85}
{Yeh}, J.~J., \& {Lindau}, I. 1985, Atomic Data and Nuclear Data Tables, 32, 1,
  \dodoi{10.1016/0092-640X(85)90016-6}

\bibitem[{{Zhang} {et~al.}(2007){Zhang}, {Richardson}, {Webb}, {Gopalswamy},
  {Huttunen}, {Kasper}, {Nitta}, {Poomvises}, {Thompson}, {Wu}, {Yashiro}, \&
  {Zhukov}}]{zhangjie07}
{Zhang}, J., {Richardson}, I.~G., {Webb}, D.~F., {et~al.} 2007, Journal of
  Geophysical Research (Space Physics), 112, A10102,
  \dodoi{10.1029/2007JA012321}

\bibitem[{{Zhao} {et~al.}(2017{\natexlab{a}}){Zhao}, {Landi}, {Lepri},
  {Gilbert}, {Zurbuchen}, {Fisk}, \& {Raines}}]{zhaoliang17a}
{Zhao}, L., {Landi}, E., {Lepri}, S.~T., {et~al.} 2017{\natexlab{a}}, \apj,
  846, 135, \dodoi{10.3847/1538-4357/aa850c}

\bibitem[{{Zhao} {et~al.}(2017{\natexlab{b}}){Zhao}, {Landi}, {Lepri},
  {Kocher}, {Zurbuchen}, {Fisk}, \& {Raines}}]{zhaoliang17b}
---. 2017{\natexlab{b}}, \apjs, 228, 4, \dodoi{10.3847/1538-4365/228/1/4}

\bibitem[{{Zhao} {et~al.}(2014){Zhao}, {Landi}, {Zurbuchen}, {Fisk}, \&
  {Lepri}}]{zhaoliang14}
{Zhao}, L., {Landi}, E., {Zurbuchen}, T.~H., {Fisk}, L.~A., \& {Lepri}, S.~T.
  2014, \apj, 793, 44, \dodoi{10.1088/0004-637X/793/1/44}

\bibitem[{{Zurbuchen} {et~al.}(2016){Zurbuchen}, {Weberg}, {von Steiger},
  {Mewaldt}, {Lepri}, \& {Antiochos}}]{zurbuchen16}
{Zurbuchen}, T.~H., {Weberg}, M., {von Steiger}, R., {et~al.} 2016, \apj, 826,
  10, \dodoi{10.3847/0004-637X/826/1/10}

\end{thebibliography}
\end{document}